%% file: template.tex
\documentclass[a4paper,twoside,10pt]{article}
\usepackage{graphicx,publaob}
\usepackage{xspace}
\usepackage{makecell,multirow}
\usepackage{titlesec}

\usepackage{natbib}

\def\papertype{Invited Lecture}
\newcommand{\rotsup}{\ensuremath{\lambda_{R_\mathrm{e}}^\mathrm{N}}\xspace}
\newcommand{\rhoio}{\ensuremath{\rho_{10}}\xspace}
\newcommand{\mjam}{\ensuremath{M_\mathrm{JAM}}\xspace}

\ExplSyntaxOn
\cs_new:Npn \expandableinput #1
  { \use:c { @@input } { \file_full_name:n {#1} } }
\AddToHook{env/tabular/begin}
  { \cs_set_eq:NN \input \expandableinput }
\ExplSyntaxOff

\let\OLDthebibliography\thebibliography
\renewcommand\thebibliography[1]{
  \OLDthebibliography{#1}
  \setlength{\parskip}{0pt}
  \setlength{\itemsep}{0pt plus 0.3ex}
}

\titleformat*{\section}{\vspace{-4ex} \center \normalsize \bfseries}
\titleformat*{\subsection}{\vspace{-4ex} \center \normalsize \bfseries}

\begin{document}

\title{\titlefont WHY DO DIFFERENT EARLY-TYPE-GALAXIES HAVE DIFFERENT AMOUNTS OF ROTATIONAL SUPPORT?}

\authors{M. B\'{i}lek$^{1,2,3,4}$,
P.-A. Duc$^3$ \lowercase{and}
E. Sola$^{3,5}$}

\address{$^1$LERMA, Observatoire de Paris, CNRS, PSL Univ., Sorbonne Univ., 75014 Paris, France
    }
\Email{michal.bilek}{obspm}{fr}

\vskip-3mm

\address{$^2$Coll\`ege de France, 11 place Marcelin Berthelot, 75005 Paris, France}

\vskip-3mm

\address{$^3$Universit\'e de Strasbourg, CNRS, Observatoire astronomique de Strasbourg (ObAS), UMR 7550, 67000 Strasbourg, France}
\vskip-3mm

\address{$^4$FZU – Institute of Physics of the Czech Academy of Sciences, Na Slovance 1999/2, Prague 182 21, Czech Republic}
\vskip-3mm
\address{$^5$ Institute of Astronomy, Madingley Road, Cambridge CB3 0HA, UK}

\markboth{\runningfont ORIGIN OF THE DIVERSITY OF ROTATIONAL SUPPORT OF ETGS}
{\runningfont M. B{\' I}LEK \runningit{et al.}}

\vskip-1mm

\abstract{Early-type galaxies (ETGs, i.e. elliptical and lenticular galaxies) differ in their amount of rotational support -- some are purely supported by velocity dispersion, while others show pronounced ordered rotation. Cosmological hydrodynamical simulations show that the progenitors of all ETGs were first rotating quickly, but then mergers decreased their rotational support. In the presented work, we studied this process using an observational archaeological approach. Namely, we inspected the correlations of 23 merger-sensitive characteristics of local ETGs with a parameter quantifying the rotational support. We used a volume-limited sample of local ETGs, that are not in galaxy clusters, from the MATLAS survey. We found, for example, that slowly rotating galaxies have tidal features and kinematically distinct components  more often and have lower metallicities. We sought for mutual interpretation of the correlations among all 23 quantities, together with literature results on high-redshift massive galaxies. There seems to be only one interpretation  possible: on average, ETGs lose their rotational support through multiple minor wet mergers happening at the redshifts above about two. 

}

\vskip-.5cm

\section{ INTRODUCTION}
\vspace{-2ex}

Galaxies are divided into early-type and late-type galaxies according to their photometric morphology. In this paper, we define early-type galaxies (ETGs) as those whose optical images do not show spiral arms and star-forming regions. In the Hubble classification, ETGs correspond to elliptical and lenticular galaxies. They appear rather featureless in images, containing either no features at all or just bars and rings. However, they turned out to be a surprisingly diverse group once they were observed with integrated field spectrographs (IFSs) \citep[e.g.,][]{emsellem11}. Their maps of line-of-sight velocities show that they have various degrees of rotational support. The so-called slow rotators show no substantial overall rotation, while  fast rotators show ordered disk-like rotation in the kinematic maps. Slow rotators are usually spherical or triaxial in shape, can contain kinematic substructures, often live in the centers of galaxy  groups and clusters and usually are more massive than fast rotators \citep{cappellari16}. Fast rotators are flattened ellipsoids, and tend to live in the field or are the non-central galaxies of their environments. This brings up the question of what caused that  different ETGs have different amounts of rotational support. We aimed to find the answer in the paper \citet{bil23} (B23 hereafter) which we summarize here.

\vskip-3mm

\section{MAIN METHOD: MULTI-TRACER ARCHAEOLOGY}
\vspace{-2ex}

We first had a look at what cosmological hydrodynamical simulation tell us about that. It turned out that all Illustris \citep{penoyre17}, EAGLE \citep{lagos22} and Magneticum \citep{schulze18} tell the same: the progenitors of all ETGs were initially quickly rotating galaxies some of which decreased the rotation later. The decrease of rotation happens usually because of mergers. For our work, we assumed that this is true and our main goal was find more about the mergers observationally, namely whether the mergers were wet (gas-rich) or dry (gas poor), minor (high mass ratio) or major (comparably massive galaxies are merging), and when they were happening.  For the galaxies with a high rotational support, we assumed that a lower fraction of their stars was accreted in mergers.

If we want to investigate the past mergers that galaxies experienced, we have to investigate the properties of the galaxies that are expected to be sensitive to mergers. In total, we investigated 23 such merger-sensitive parameters, such as the presence of tidal features, kinematically distinct cores, chemical composition of the galaxies, etc. Each of these parameters reacts differently to different types of mergers and has a different lifetime. The investigation of that many parameters at once turned out to be crucial, since this enabled breaking degeneracies in the possible interpretation. 

We took the data from the intersection of the MATLAS \citep{bil20} and ATLAS$^\mathrm{3D}$ \citep{cappellari11a} databases. ATLAS$^\mathrm{3D}$ is a complete volume-limited survey of nearby ETGs. It was originally designed as an IFS survey, but many properties of its targets have been published. MATLAS took images reaching a very low surface-brightness limit for the ATLAS$^\mathrm{3D}$ galaxies that are outside of galaxy clusters. Its aim was  to detect tidal features in the galaxies, i.e. morphological disturbances caused by past mergers. Our sample included 175 galaxies in total, with stellar masses between about $10^{10}$ and $10^{12}\,M_\odot$. Literature findings on high-redshift galaxies helped us to interpret the results.

For investigating the process of decreasing of the rotational support of galaxies, it is necessary to quantify the rotational support. We used the \rotsup parameter introduced by \citet{emsellem11}. Roughly speaking, \rotsup depends on the rotational velocity of the galaxy, its velocity dispersion, and ellipticity. A higher value of \rotsup indicates a higher rotational support.

One might be tempted to inspect the correlations with \rotsup with the merger-sensitive parameters and to use them to deduce the properties of the mergers that were responsible for the decrease of the rotational support. However, we caution that such an approach could give misleading results. Let us explain why on an analogy. Suppose that we want to investigate the impact of smoking on the health of people. However, in the sample of people we have at disposal, the smoker are heavier than the non-smokers. Thus, if we detect that smoking correlates with some measure of health, it would be not clear whether the correlation is a result of the smoking or of the obesity. The solution of this is to investigate the impact of smoking on health for people of a fixed weight. 

The situation is similar for galaxies, because the less rotating galaxies tend to be more massive and live in denser environments. At the same time, the strength of many processes shaping galaxies, different from mergers, depend on these quantities too.  For example, the metallicity of a galaxy, which was one of the merger-sensitive parameters we used, is influenced by the ability of the galaxy to retain the gas enriched by metals by the stellar evolution. If the galaxy is more massive, the winds from supernovae and active galactic nuclei have a harder job to remove the metals from the galaxies. Likewise, the removal of gas by ram-pressure stripping is more important for denser environments. We must therefore investigate the correlations between merger-sensitive parameters and the rotational support at fixed mass and environmental density.

To ensure this, we used the method known as multilinear regression. Namely, we fitted the value of a merger-sensitive parameter by a linear combination of the logarithm of the stellar mass derived from Jeans anisotropic modelling \citep{cappellari08}, \mjam, of environmental density, \rhoio, and rotational support, \rotsup, as: 
\begin{equation}
    [\textrm{parameter}] = b+a_M\log\mjam +a_\rho \log\rhoio +a_\mathrm{KS}\rotsup.
    \label{eq:fit}
\end{equation}
In such a fit, the coefficient at \rotsup, $a_\mathrm{KS}$, is the correlation coefficient between the merger-sensitive parameter and \rotsup at fixed mass and environmental density. Next, it was necessary to evaluate the statistical significance of the correlation, i.e. the probability that the detected correlation is there just as a coincidence. To this point, we repeated the same fit as before, but without the last term, $a_\mathrm{KS}\rotsup$. An $F$-test applied on the residuals of the two fits indicated whether the last term holds any statistically significant information about the merger-sensitive parameter.

Table~\ref{tab:tab1} shows the result. The first column lists the different considered merger-sensitive parameters (see their explanations below). The second column indicates the result when the test described above is applied to the whole sample of 175 galaxies. The zero sign indicates that the correlation is not statistically significant at the 5\% confidence level. The plus or minus signs mean that the given correlation is statistically significant and is positive or negative, respectively. The N/A symbol indicates that there were too few galaxies to calculate the result. In the last four columns, we separated the sample into two parts, either by the \mjam mass (in solar masses)  or by the environmental density \rhoio (units of Mpc$^{-3}$). 

Now we briefly explain the different merger-sensitive parameters in the table (see B23 for details). Shells, streams and tails are tidal features of specific shapes. Disturbed isophotes are just minor photometric irregularities caused probably by old or weak  interactions. The next two parameters  indicate whether the galaxy has any tidal features, and any tidal features or disturbed isophotes, respectively. The quantities Shell, stream and tail frequency indicate the number of these features in the particular galaxy. The next parameter indicates the presence of dust lanes. Kinematically distinct cores, counterrotating cores, kinematic twists and the $2\,\sigma$ structures are specific features seen in the kinematic maps. $R_e$ indicates the value of effective radius of the galaxy and $\gamma^\prime$ the value of the slope of the radial photometric profile in the galactic center. The last quantities are the average stellar ages, metallicities, and alpha-element abundances in the galaxies,  and their radial gradients from the single stellar population models of the galaxies.

\begin{table}[t!]
\caption{Correlations between the different merger-sensitive parameters and  the rotational support \rotsup at fixed mass and environment density.}
\vskip2mm
\centerline{\begin{tabular}{l|ccccc}
		\hline
 \multirow{3}{*}{Parameter} & \multicolumn{5}{c}{Sample}\\ \cline{2-6}
 & \makecell[c]{All\\galaxies} & \makecell[c]{$\log \mjam$\\$<11$}  & \makecell[c]{$\log \mjam$\\$\geq11$} & \makecell[c]{$\log \rhoio$\\$<-2$} & \makecell[c]{$\log \rhoio$\\$\geq-2$}\\
\hline
\input{Ftest_rotmes_tar-all-proceedings.txt}
\hline
	\end{tabular}}
    \label{tab:tab1}
\end{table}


\subsection{INTERPRETATION}
\vspace{-2ex}

We found only one way to explain the above correlations simultaneously: ETGs were primarily losing their rotational support when they still contained a substantial fraction of gas, that is at the redshifts higher than about two (i.e., 10\,Gyr ago), because of numerous wet minor mergers. Moreover, the galaxies with a low rotational support tend to experience more mergers until today than the galaxies with a higher rotational support. These two sentences contain five claims, which we justify below. 

\textit{1) The rotational support of ETGs was decreased by mergers.} This was an assumption of our work inspired by simulations, but there indeed is observational evidence for it. We tried hard to find an explanation of the correlations in Table~\ref{tab:tab1} without mergers, but we did not find any. It is possible to explain every single correlation without mergers (even if the explanation might not appear fully convincing), but the real difficulty is in explaining the correlations \textit{simultaneously}. For example, we found that at fixed mass and environment density, galaxies with a lower rotational support have a lower metallicity. It is observed that less massive galaxies generally have lower metallicities than more massive galaxies. Therefore it looks plausible that the galaxies with a low rotational support inherited the low metalicities of their progenitors. 

\textit{2) The mergers that decreased the rotational support were wet.} We have three independent pieces of evidence for this. First, we found for galaxies at fixed mass and environment density, that the galaxies with a low rotational support have steeper central slopes of the radial photometric profiles. For some galaxies, the slope becomes more shallow toward the center, such that the profile has a core. For others, the profile remains steep toward the center and then we speak about a cuspy profile. The central photometric profile reacts to mergers according to the amount of available gas. Let us consider first a dry (gasless) merger. Each has a supermassive black hole in its center. After the galactic merger, the two black holes give their orbital energy and angular momentum to the surrounding stars, such that they settle in the center of the galaxy until they merge themselves. During this process, they eject stars from the galaxy center, such that it becomes deficient in stars and a photometric core forms. On the other hand, if there is gas available, it concentrates toward the center of the resulting galaxy, and forms new stars. The deficit of stars gets replenished, such that the galaxy can even become more cuspy than before the merger. This is what we observe for the galaxies with a low rotational support.

The second piece of evidence for wet mergers is that for galaxies at a fixed mass and rotational support, we found that the less rotating galaxies have a higher abundance of alpha elements. A high abundance of alpha elements indicates an intensive star formation. It is observed that the alpha element abundance increases with the mass of the galaxy. If the mergers that formed the slowly rotating galaxies were dry, the galaxies would simply inherit the low alpha-element abundances of their progenitors. However, we instead observe that the less rotating galaxies have increased alpha element abundances. This indicates that the mergers that created them had to be gas-rich, such that they could give rise to intensive starbursts.

Finally, we found that for galaxies at a fixed mass and environment density, there is no significant correlation between the rotational support and the effective radius of the galaxy. If we have a dry merger of two galaxies, their initial relative potential and kinetic energies have to transform into the internal energy of the merger remnant. Consequently, the resulting galaxy has to be larger than the progenitors. But we do not observe that the galaxies with a low rotational support have larger radii. This indicates that the energy has to be removed. This can be achieved by the merger progenitors being gas rich, such that the merger energy was  removed by radiation. 

\textit{3) The mergers that decreased the rotational support were minor.} We have just one piece of evidence for this, such that this is the least robust result. Namely, we found that for galaxies at a fixed mass and environment density, the galaxies with a lower rotational support have steeper metallicity gradients. To interpret this, we first notice that in most galaxies, the metallicity decreases from the center toward the edge. In a major merger of two galaxies, regardless of their gas content, the resulting metallicity gradient is lower than it was initially, simply because of the mixing of the stellar populations. This is not what we observe for the galaxies with a low rotational support.  The observed correlation can only be explained by the mergers being minor. For them, it is known that the material of the metal-poor smaller galaxy tends to be deposited at large galactocentric radii of the bigger galaxy, such that the metallicity gradient becomes steeper. 

\textit{4) The mergers that decreased the rotational support were happening at a high redshift.} The first piece of evidence  comes simply from that the mergers that formed the ETGs with a low rotational support were wet, and that ETGs typically stopped forming stars at the redshift of about two, i.e. 10\,Gyr ago (while this depends on the mass of the galaxy and probably other details).

Second, there is no statistically significant correlation between the amount of rotational support and stellar age for galaxies at a fixed mass and environment density. Late wet mergers would lead to intensive starbursts and decreasing of the average stellar age. This can be explained by the mergers happening a long time ago:  if we have two stellar populations with slightly different ages, then it is becoming increasingly difficult to detect the difference in the ages the older the two populations are.

\begin{figure}
	\centering
\includegraphics[width=0.75\textwidth,keepaspectratio=true]{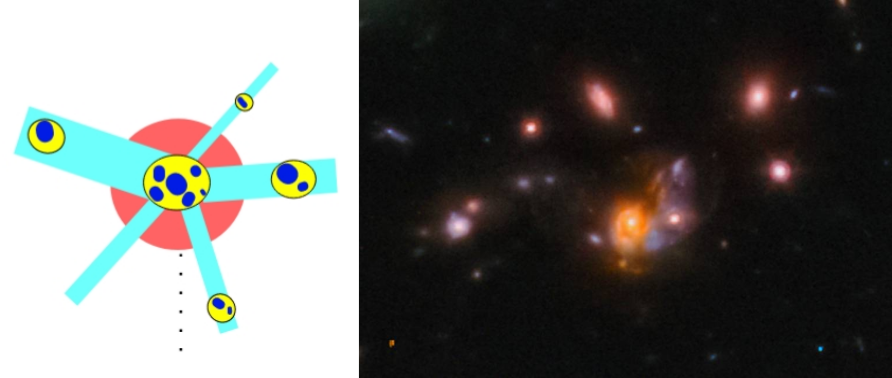}

\vspace{-3mm}

\caption{\emph{Left}: Cartoon of the formation of a slow rotator as deduced from the properties of the local ETGs, i.e. the galaxy forms from multiple minor gas-rich mergers at about $z=2$ (from B23).  \emph{Right}: Similar scenes are now seen by JWST. \citet{coogan23} described this compact group of 16 interacting starforming galaxies at $z=1.85$ (from the CEERS public image). The central galaxies of groups are known to often be slow rotators.}
	\label{fig:jwst}
\vspace{-1mm}
\end{figure}

Then we have numerous indications of the early formation of the galaxies with a low rotational support from the observations of galaxies at high redshifts. It is known that most of the galaxies with masses over $10^{11}\,M_\odot$ are slow rotators \citep{emsellem11}. If we count the volume density of such galaxies (i.e., the number of galaxies per cubic megaparsec) as a function of redshift, we find that it virtually stopped increasing 9\,Gyr ago \citep{kawinwanichakij20}. The most massive passive galaxies have always had a low on-sky ellipticity for the last at least 10\,Gyr \citep{chang13}, which indicates no overall rotation, which would make the galaxies flattened, and have had high S\'ersic indexes, typical for non-rotating galaxies, since at least 12\,Gyr ago \citep{lustig21}. Moreover, if the stellar populations of massive quenched galaxies observed at the look-back time of 9.6\,Gyr ago evolved passively, their metallicity and alpha element abundances would match excellently those of the local ETGs \citep{onodera15}. Brightest cluster galaxies, which often are slow rotator ETGs in the nearby Universe \citep{jimmy13,vandesande21b}, are observed to have non-evolving luminosities, surface brightnesses and effective radii at least to the look-back time of 10\,Gyr \citep{chu21,chu22}. The recent images by the James Webb Space Telescope are indeed suggestive of our scenario, see Fig.~\ref{fig:jwst}. 

\textit{5) Galaxies with a low rotational support keep experiencing more mergers until today.} We think so, because we found for galaxies at a fixed mass and environmental density, that the galaxies with a lower rotational support contain disturbed outer isophotes more often. There is also a tendency for them, yet insignificant, to host tidal features more often.  At the same time, the lifetime of these features is estimated to be at most 4-9\,Gyr.  Therefore the observed disturbed isophotes and tidal features cannot be remnants of the old mergers which we identified before to be responsible for the decrease in the rotational support.

\vskip-3mm

\section{ESTIMATE OF MERGING RATE OF GALAXIES FROM THE FREQUENCY OF TIDAL DISTURBANCES}
\vspace{-2ex}

In the previous section, we found that the ancient mergers were mainly responsible for decreasing the rotational support of ETGs, even if the little rotating galaxies are still experiencing more mergers now. For a sanity check, we wanted to convince ourself again that the late mergers are not too important in the evolution of ETGs. We thus estimated the recent merging frequency of ETGs from the fraction $f$ of galaxies that have tidal features or at least disturbed outer isophotes. In B23, we derived a formula to estimate the average number of interactions that have created the currently observed tidal disturbances:
\begin{equation}
     n = -\ln(1-f).
     \label{eq:n}
\end{equation}
The fraction $f$ was determined from the very deep MATLAS images of the galaxies \citep{bil20}. If we count all galaxies that have at least a possible hint of a tidal disturbance and assume the shortest possible lifetime of tidal disturbances and that all the tidal disturbances were formed by mergers, we still get a very low current merging frequency of 2 mergers per ten gigayears. This includes even very minor mergers: most tidal features detected in MATLAS contain just 1\% of the total luminosity of the galaxy (Sola et al., in prep.), and thus we can detect past mergers of the stellar mass ratio of perhaps even 1:100. We then selected galaxies that appear to have undergone a major merger according to their disturbances.  Using Equation~\ref{eq:n} and the shortest possible lifetime of the disturbances, we estimated the frequency of major mergers to be just 0.2 per ten gigayears. These estimates of merging frequencies agree with the literature observational estimates using different methods. The late mergers thus do not seem important for the evolution of ETGs.

This has an interesting consequence. It is observed that high-redshift quenched galaxies have smaller sizes than the local ones with the same stellar mass. The expansion in radius is usually accounted to mergers. \citet{trujillo06} calculated how many mergers are needed since  $z=0.4$. The actual merging frequency we found is a few times lower, especially for the lighter of our galaxies. Therefore there must be some additional mechanism causing the expansion of ETGs. 

\vskip-3mm

\section{CONCLUSIONS: FORMATION OF MASSIVE GALAXIES}
\vspace{-2ex}

We can put the above findings together with other results from the literature to build a picture of formation of typical ETGs with different amounts of rotational support and of massive spirals. Let us begin with the formation of a typical slow rotator. At redshifts higher than about two, there are strong inflows of intergalactic gas in the galaxy \citep{dekel06}. At the same time, it experiences the multiple minor mergers whose signs we detected. At the redshift of around one, once the galaxy crosses a certain mass limit, the inflow of the intergalactic gas cannot feed star formation in the galaxy any longer, because it is shock-heated so much that the cooling time exceed the age of the Universe  \citep{dekel06}. Instead, it forms a hot atmosphere around the galaxy. As cosmic time proceeds, the merging frequency decreases. The mergers are becoming dry, in part because galaxies generally become less gas rich at later redshifts,  and moreover because the galaxies to be accreted are cut off from the inflow of cold intergalactic gas when they move in the hot atmosphere of the main galaxy. At the redshift of zero, mergers are very rare.

The formation of a spiral seems to be an opposite of the formation of a slow rotator. It is observed that most spiral galaxies do not have classical bulges \citep{kormendy10} and do not possess substantial stellar halos \citep{merritt16}, structures that form in mergers. For the Milky Way, it was found that only a few percent of its stars were accreted from other galaxies \citep{helmi20,forbes20}. Therefore, if any substantial growth by mergers was happening, it had to be when the mergers were very gas rich at high redshifts. The star formation rate of the Milky Way has been almost constant in the last 9\,Myr, when it formed most of its stars \citep{fantin19}, such that the galaxy grows rather gradually.

The fast rotator ETGs appear to be an interpolation between spirals and slow rotators: their mass assembly is faster than of spirals but not as fast as that of slow rotators. They experience more mergers than spirals but less than slow rotators. 

One should however remember that these were just the typical evolutionary paths of the three galaxy types. Some galaxies can deviate from them. Also, we caution again that these results were derived for ETGs exclusively outside of galaxy clusters.

\vskip2mm




\vskip-.5cm


\bibliographystyle{aa-short}
\bibliography{tmp.bib}


\end{document}

%% file: Ftest_rotmes_tar-all-proceedings.txt
Shells              & 0& 0& 0& 0& 0 \\
Streams             & 0& 0& 0& 0& 0 \\
Tails               & 0& 0& 0& 0& 0 \\
Disturbed isophotes & -& 0& 0& 0& - \\
Any TF              & 0& 0& 0& 0& 0 \\
Any TF or DI        & 0& 0& 0& 0& 0 \\
Shell fr.           & 0& 0& 0& 0& 0 \\
Stream fr.          & 0& 0& 0& 0& 0 \\
Tail fr.            & 0& 0& 0& 0& 0 \\
Dust                & 0& 0& 0& 0& 0 \\
KDC                 & -& -& 0& 0& - \\
CRC                 & -& -& 0& -& - \\
KT                  & 0& 0& 0& N/A & 0 \\
$2\,\sigma$         & -& -& N/A & 0& - \\
No kin. feature     & +& +& 0& +& 0 \\
$\log R_e$          & 0& 0& 0& 0& 0 \\
Core $\gamma^\prime$& 0& 0& 0& -& 0 \\
SSP age             & 0& 0& 0& 0& 0 \\
SSP age gr.         & 0& 0& 0& 0& 0 \\
SSP Z               & +& +& 0& +& + \\
SSP Z gr.           & +& 0& 0& +& 0 \\
SSP $\alpha$        & 0& 0& -& 0& 0 \\
SSP $\alpha$ gr.    & 0& 0& 0& 0& 0 \\